# First single crystal growth and structural analysis of superconducting layered bismuth oxyselenide; La(O,F)BiSe$_2$

Masashi TANAKA[1*], Masanori NAGAO[2], Yoshitaka MATSUSHITA[1], Masaya FUJIOKA[1], Saleem J. Denholme[1], Takahide YAMAGUCHI[1], Hiroyuki TAKEYA[1], and Yoshihiko TAKANO[1]

[1]National Institute for Materials Science, 1-2-1 Sengen, Tsukuba, Ibaraki 305-0047, Japan

[2]University of Yamanashi, 7-32 Miyamae, Kofu, Yamanashi 400-8511, Japan

*Corresponding author:

Masashi Tanaka

E-mail: Tanaka.Masashi@nims.go.jp

Postal address: National Institute for Materials Science

1-2-1 Sengen, Tsukuba, Ibaraki 305-0047, Japan

**Abstract**

Single crystal of La(O,F)BiSe$_2$ was successfully grown for the first time by the CsCl flux method. Single crystal X-ray structural analysis clearly showed that La(O,F)BiSe$_2$ crystalizes with space group *P4/nmm* (lattice parameters $a$ = 4.1408(2) Å, $c$ = 14.1096(6) Å). The structure is composed of alternating BiSe$_2$ and LaO layers. Magnetic susceptibility and resistivity measurements showed the superconducting transition at 3.6 K, which is higher than that of as-grown La(O,F)BiS$_2$. The structural analysis implies that La(O,F)BiSe$_2$ is potential superconductor having higher $T_c$ under high pressure.

**1. Introduction**

After a discovery of superconductivity in Bi$_4$O$_4$S$_3$ [1], much attention has been paid for synthesizing the BiS$_2$-based compounds. The BiS$_2$-based compounds have a layered structure composed of superconducting BiS$_2$ layers and charge reservoir blocking layers similar to cuprate or Fe-based superconductors. For the Bi$_4$O$_4$S$_3$ case, it is composed of a superconducting layer Bi$_2$S$_4$ and a blocking layer Bi$_4$O$_4$(SO$_4$)$_{1-x}$. There are BiS$_2$-based superconductors having another type of the blocking layer LaO, which is similar to that of Fe-based superconductor LaFeAsO [2]. La(O,F)BiS$_2$ synthesized under ambient pressure shows superconductivity with a transition temperature ($T_c$) of ~2.5 K [3]. The $T_c$ is enhanced as high as 11.5 K under high pressure [4], and this is the highest record of BiS$_2$-based superconductors. BiS$_2$-based superconductors with various types of blocking layer have been developed up to date, such as $Ln$(O,F)BiS$_2$ ($Ln$ = La, Ce, Pr, Nd, Yb) [3, 5-9], (La,$M$)OBiS$_2$ ($M$ = Th, Hf, Zr, Ti) [10], (Sr,La)FBiS$_2$ [11]. $Ln$(O,F)BiS$_2$ ($Ln$ = La, Ce, Nd) have been prepared as single crystals using alkali metal chloride flux methods [12-14]. The single crystal X-ray structural analysis for La(O,F)BiS$_2$ and Nd(O,F)BiS$_2$ determined the bonding distances and angles precisely [11,15]. Especially in the case of Nd(O,F)BiS$_2$, the structural investigation of different F contents were performed by Miura *et al*. [15]. They showed that the substitution of O by F results in the distortion of the Bi-S plane.

On the other hand, modification of superconducting layers is a very challenging issue. Only two cases have been reported in polycrystalline samples: LaO$_{0.5}$F$_{0.5}$BiSSe ($T_c$ ~3.8 K) [16] and La(O,F)BiSe$_2$ ($T_c$ ~2.6 K) [17]. It is necessary to obtain the single crystal of BiSe$_2$-based material in order to clarify the superconducting mechanism and precise crystal structure for design a new compound with higher $T_c$.

In this study, we have succeeded in preparing single crystal of La(O,F)BiSe$_2$ by the CsCl flux method and carried out various characteristic measurements including single crystal X-ray diffraction. Here, we discussed bonding nature and superconductivity of this compound.

## 2. Experimental

*2.1. Preparation of single crystals*

Single crystal of La(O,F)BiSe$_2$ was prepared by a flux method using CsCl. The powders of La (Kojundo, 99.9%), Bi$_2$O$_3$ (Wako, 99.9%), BiF$_3$ (Kojundo, 99.9%), Bi (Kojundo, 99.99%), Bi$_2$Se$_3$ (Kojundo, 99.9%), Se (Kojundo, 99.9%) were mixed in nominal compositions of LaO$_{0.5}$F$_{0.5}$BiSe$_2$. The mixed powders with 5 g of CsCl (Kojundo, 99.9%) were sealed in an evacuated quarts tube and heated to 880 °C in 10 h, then gradually heated up to 900 °C in 2h. The temperature was kept for 10 h, followed by cooling to 650 °C at a rate of 1 °C/h, then the sample was cooled down to room temperature in the furnace. The obtained materials were washed and rinsed by water, ethanol, and acetone in order to remove the flux material CsCl.

*2.2. Characterization*

Single crystal X-ray structural analysis was carried out using a Rigaku AFC11 Saturn CCD diffractometer with a VariMax confocal X-ray optical system for Mo K$\alpha$ radiation ($\lambda$ = 0.71073 Å). Prior to the diffraction experiment, the crystals were cooled to 160 K in a cold N$_2$ gas flow, in order to suppress thermal displacement factor. Cell refinement and data reduction were carried out by the program d*trek package in CrystalClear software suite [18]. The structure was solved by the direct method using SHELXS [19], and was refined with the program SHELXL [19] in WinGX software package [20]. Back scattered electron (BSE) images and energy-dispersive X-ray (EDX) spectra were observed using a scanning electron microscope (Hitachi SU-70, EDAX Genesis Apollo XP).

Temperature dependence of magnetic susceptibility was measured using a SQUID magnetometer (MPMS, Quantum Design) down to 2.0 K under a field of 10 Oe, and the field was applied parallel to the *c*-axis. Temperature dependence of electrical resistivity was measured down to 2.0 K, using a standard four-probe method with constant current mode by a Physical Property Measurement System (PPMS, Quantum Design). The electrodes were attached in the *ab*-plane with silver paste.

## 3. Results and Discussion

*3.1 Crystal structure of the La(O,F)BiSe$_2$*

BSE image of the typical single crystal is shown in Fig.1(a). The image suggests the crystal has a uniform composition, and no local precipitation of metal. The composition was analyzed by EDX point analysis at several sections of the crystal. Fig.1(b) shows a typical EDX spectrum at a point marked by the red cross in Fig.1(a). The averaged compositional ratio of elements is estimated to be La : Bi : Se : O : F = 0.98 : 0.92 : 2 : 0.50 : 0.19, in which the ratio was normalized to Se = 2. The atomic ratio except for O and F elements is in good agreement with the nominal composition, suggesting that the single crystal has a composition La(O$_{1-x}$F$_x$)BiSe$_2$. It should be noted that the analysis always detected a slight excess of La in relation to Bi.

The X-ray single crystal analysis was successfully performed. Details of the single-crystal structure analysis at 160 K and crystallographic parameters are listed in Table 1 and Table 2. The compound crystallizes with space group *P4/nmm* (lattice parameters $a$ = 4.1408(2) Å, $c$ = 14.1096(6) Å, and $Z$ = 1). All atoms were located on special position and their anisotropic displacement factors were all positive with the similar values, appearing physically reasonable. The basic structure of this compound showed isostructure with La(O,F)BiS$_2$ [3] and previously reported polycrystalline La(O,F)BiSe$_2$ [17]. The ratio of O and F was estimated to be O : F = 0.82 : 0.18, namely La(O$_{0.82}$F$_{0.18}$)BiSe$_2$, from the occupancy refinement under constrain of Occ.(O) + Occ.(F) = 1. The F content $x$ is in good agreement with the F content of EDX result of $x$ = 0.19.

A schematic illustration of the crystal structure is shown in Fig. 2(a). The crystal has an alternate stacking of $BiSe_2$ pyramids and $La_2(O/F)_2$ layer similar to $BiS_2$-based superconductors. The ORTEP representation is shown in Fig. 2(b). It is necessary to mention about residual peaks/holes around Bi and La sites. There are relatively large residual maximum densities ($<\sim5$ $e^-/Å^3$) around Bi and La sites even though the $R1$ value falls as small as ~3 %. Fig. 3 shows the precession photographs reconstructed from the observed single crystal diffraction data. It is clearly shown that no modulated structure was observed. This finding suggests that the large residual densities are attributed to local disorder of Bi/La-sites in its crystal structure similar to X-ray absorption spectroscopy results in $BiS_2$-based materials [21]. Since the EDX result always shows La-rich content compare than that of Bi, one of a possibility is that a little amount of La migrates into Bi site as shown in Fig. 2(b) (Bi' and La' sites). In this model, the total occupancy and valence of Bi and La were constrained to keep 1.0 and +3, respectively. The final refined composition was $La_{1.010}(O_{0.82}F_{0.18})Bi_{0.992}Se_2$. However, we cannot exclude the possibility of self-disorder of Bi atoms. There are no considerable differences in $R1$ values between the case of Bi self-disorder and Bi/La disorder, although the lowest $R1$ value was obtained in the model described above.

Some selected bond distances and angles are given in Fig. 2(b) and Table 3. Bi atoms are six coordinated and bonded with Se atoms in three kinds such as Bi-Se1, intra-plane Bi-Se2, and inter-plane Bi-Se2 bondings. The interatomic distance of intra-plane Bi-Se2 is 2.9317(1) Å. On the other hand, the inter-plane Bi-Se2 is much longer (3.3242(9) Å) and the Bi-Se1 is shorter (2.6752(8)

Å) than that of the intra-plane Bi-Se2. This tendency of bond lengths is similar to that of La(O,F)BiS$_2$ [3]. The intra-plane Se-Bi-Se angle in the present crystal is 174.262(15) °. In the La(O,F)BiS$_2$, it has been reported that the amount of F affects not only carrier concentration but also a distortion of the Bi-S plane [14]. The Bi-Se plane is more distorted than the Bi-S plane in La(O$_{1-x}$F$_x$)BiS$_2$ with $x \sim 0.23$.

Bond valence sum (BVS) was estimated from the observed bond distances and the $r_0$ and $B$ values of nominal valence La$^{3+}$, Bi$^{3+}$, Se$^{2-}$, O$^{2-}$, F$^-$ provided by Brown [22]. The BVS values are also listed in Table.2. According to the refined composition La$_{1.010}$(O$_{0.82}$F$_{0.18}$)Bi$_{0.992}$Se$_2$, the nominal valence summation should be +0.19, based on nominal valences. On the other hand, the BVS of Bi and La was estimated to be +6.11, indicating +0.11 different from charge neutrality. Both results indicate the refined structure is slightly cation rich, suggesting the charge neutrality should be kept by modification of valence state of Bi like as the result of X-ray photoemission spectroscopy [23].

*3.2 Superconductivity*

Fig. 4 shows the temperature dependence of the magnetic susceptibility for the single crystal La(O,F)BiSe$_2$. The diamagnetic signal corresponding to superconductivity appears below 3.6 K. The shielding and Meissner volume fraction at 2 K was estimated to be ~8 % and ~7 %, respectively. The inset shows the temperature dependence of the resistivity along the *ab*-plane in the single crystal. It showed a sharp resistivity drop at 3.6 K, and zero resistivity was observed at 3.2 K, in accordance with the superconducting transition temperature $T_c$ found in the magnetic susceptibility measurement.

Note that the $T_c$ is 1 K higher than that of previously reported polycrystalline result [17]. Currently, the reason of the $T_c$ difference is unclear but assuming that it is attributed to the F content and higher crystallinity. X-ray single crystal analysis showed the corrugation in the Bi-Se layer is slightly larger than the Bi-S layer of the $BiS_2$-based superconductor. A flat Bi-S plane would result in better hybridization of the $p_x/p_y$ orbitals of Bi and S [15]. If the Bi-Se plane could flatten by applying chemical or physical pressure, such as optimizing the F-content or uniaxial compression, overlapping in Bi- and Se-$p$ orbitals would be better and may realize higher $T_c$ in this system.

**4. Conclusion**

Single crystals of $La(O,F)BiSe_2$ were successfully grown by the CsCl flux method. The X-ray structural analysis revealed $La(O,F)BiSe_2$ crystalized with space group *P4/nmm* (lattice parameters *a* = 4.1408(2) Å, *c* = 14.1096(6) Å). The magnetic susceptibility and resistivity measurement clearly showed the superconducting transition at 3.6 K. The structural analysis suggested a possibility of raising up the $T_c$ if the corrugated Bi-Se layer could be flatten by applying chemical or physical pressure as well as $BiS_2$-based superconductors.

**Acknowledgment**

This work was partially supported by Japan Science and Technology Agency through Strategic International Collaborative Research Program (SICORP-EU-Japan) and Advanced Low Carbon Technology R&D Program (ALCA) of the Japan Science and Technology Agency.

Table 1. Crystallographic data for the La(O,F)BiSe$_2$

| | |
|---|---|
| Formula | La(O$_{0.82}$F$_{0.18}$)BiSe$_2$ |
| Formula weight | 522.36 |
| Crystal system | Tetragonal |
| Space group | *P4/nmm* (No.192) |
| *a* (Å) | 4.1408(2) |
| *c* (Å) | 14.1096(6) |
| *V* (Å$^3$) | 241.93(3) |
| *Z* | 1 |
| $d_{calc}$ (g/cm$^3$) | 7.174 |
| Temperature (K) | 160 |
| $\lambda$ (Å) | 0.71073 (MoK$\alpha$) |
| $\mu$ (mm$^{-1}$) | 59.900 |
| Absorption correction | multi-scan |
| $\theta_{max}$ (°) | 46.242 |
| Index ranges | -7<*h*<8, -8<*k*<8, -28<*l*<28 |
| Total reflections | 9173 |
| Unique reflections | 700 |
| Observed [$I \geq 2\sigma(I)$] | 694 |
| $R_{int}$ for all reflections | 0.0244 |
| No. of variables | 23 |
| *R*1/*wR*2 [$I \geq 2\sigma(I)$] | 0.0243/0.0634 |
| *R*1/*wR*2 (all data) | 0.0244/0.0636 |
| *GOF* on $F_o^2$ | 1.158 |
| Max./Min. residual density (e$^-$/Å$^3$) | 4.923 / -2.039 |

Table 2. Atomic coordinates, atomic displacement parameters (Å$^2$), and bond valence sum (BVS) for the La(O$_{0.82}$F$_{0.18}$)BiSe$_2$

| Site | Wyck. | S.O.F | x/a | y/b | z/c | $U_{11}$ | $U_{22}$ | $U_{33}$ | $U_{eq}$ | BVS |
|---|---|---|---|---|---|---|---|---|---|---|
| La | 2c | 1 | 1/4 | 1/4 | 0.09052(3) | 0.01474(10) | 0.01474(10) | 0.01281(14) | 0.01410(9) | +2.94 |
| Bi | 2c | 0.98 | 1/4 | 1/4 | 0.62301(2) | 0.01420(8) | 0.01420(8) | 0.01177(16) | 0.01339(8) | +3.10 |
| Bi' | 2c | 0.01 | 1/4 | 1/4 | 0.5855(18) | | | | 0.003(5) | +0.02 |
| La' | 2c | 0.01 | 1/4 | 1/4 | 0.664(2) | | | | 0.004(6) | +0.08 |
| Se1 | 2c | 1 | 1/4 | 1/4 | 0.81260(5) | 0.01343(15) | 0.01343(15) | 0.0102(2) | 0.01235(11) | -2.17 |
| Se2 | 2c | 1 | 1/4 | 1/4 | 0.38739(6) | 0.01299(16) | 0.01299(16) | 0.0190(3) | 0.01499(13) | -2.06 |
| O | 2a | 0.82 | 3/4 | 1/4 | 0 | 0.0131(11) | 0.0131(11) | 0.0120(14) | 0.0128(8) | -1.62 |
| F | 2a | 0.18 | 3/4 | 1/4 | 0 | 0.0131(11) | 0.0131(11) | 0.0120(14) | 0.0128(8) | -0.26 |

Note: $U_{12}$, $U_{13}$, and $U_{23}$ are 0, and $U_{eq}$ is defined as one third of the trace of the orthogonalized $U$ tensor.

Table 3. Selected bond lengths (Å) and angles (°) of the La(O$_{0.82}$F$_{0.18}$)BiSe$_2$ single crystal

| Distance | | | |
|---|---|---|---|
| La-O/F × 4 | 2.4327(2) | Bi-Se1 | 2.6752(8) |
| La-Se1 × 4 | 3.2314(4) | Bi-Se2 [inter plane] | 3.3245(9) |
| | | Bi-Se2 [in plane] × 4 | 2.9317(1) |
| Angle | | | |
| La-O/F-La | 105.998(8) | Se1-Bi-Se2 | 92.869(14) |
| La-Se1-La | 79.690(11) | Se2-Bi-Se2 [inter-plane] | 87.131(13) |
| | | Se2-Bi-Se2 [intra-plane] | 174.262(15) |

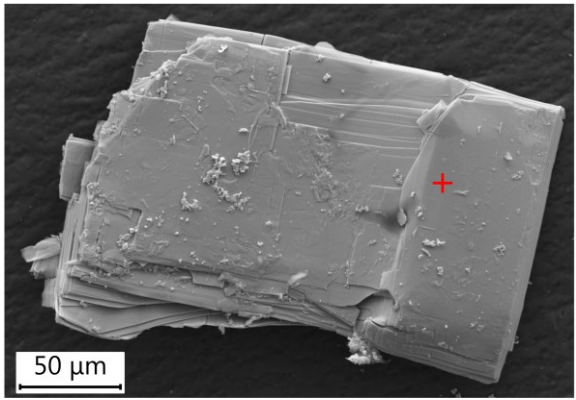 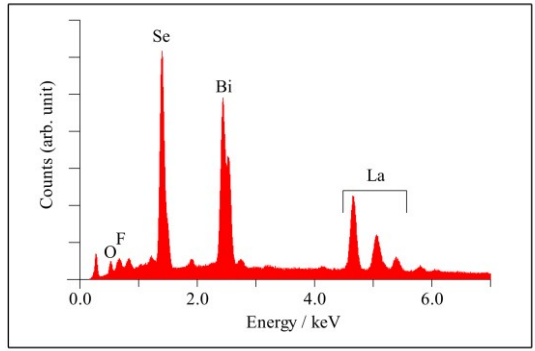

Figure 1. BSE image (a) and EDX result (b) of the obtained single crystal.

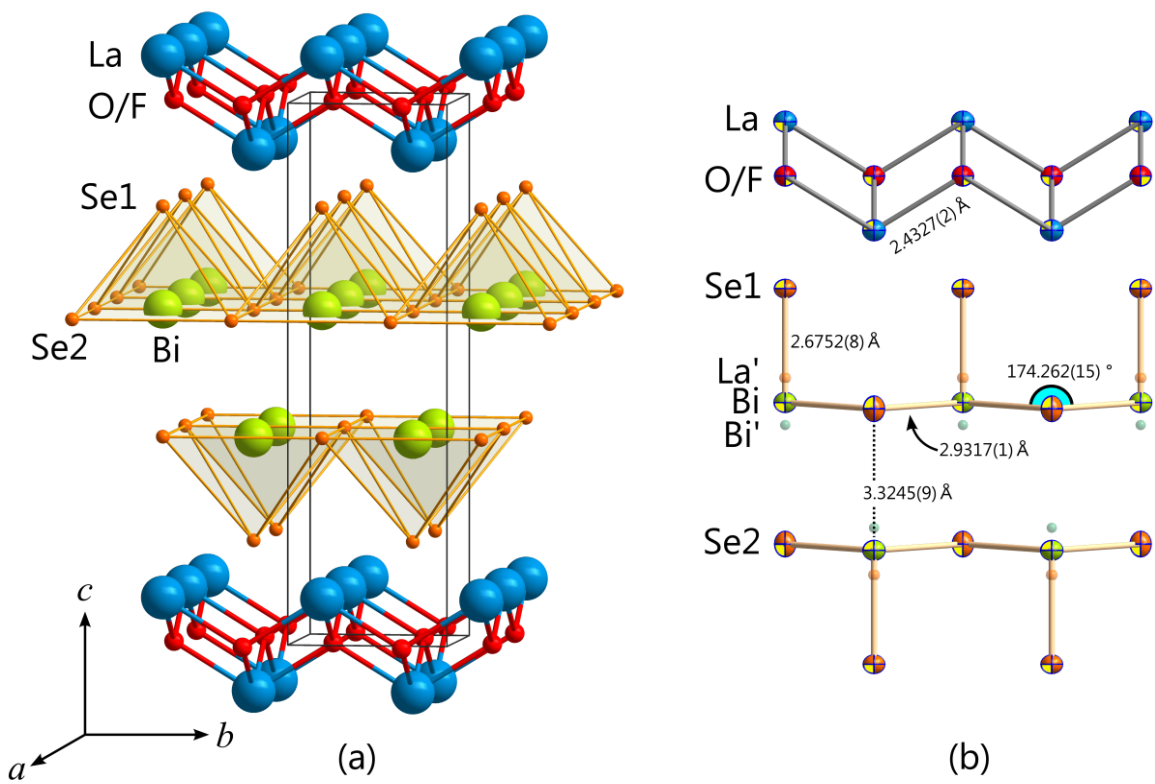

Figure 2. Crystal structure of La(O,F)BiSe$_2$ (a) and its local ORTEP representation with the selected bonding distances and angles (b). Displacement ellipsoids are drawn at the 80 % probability level.

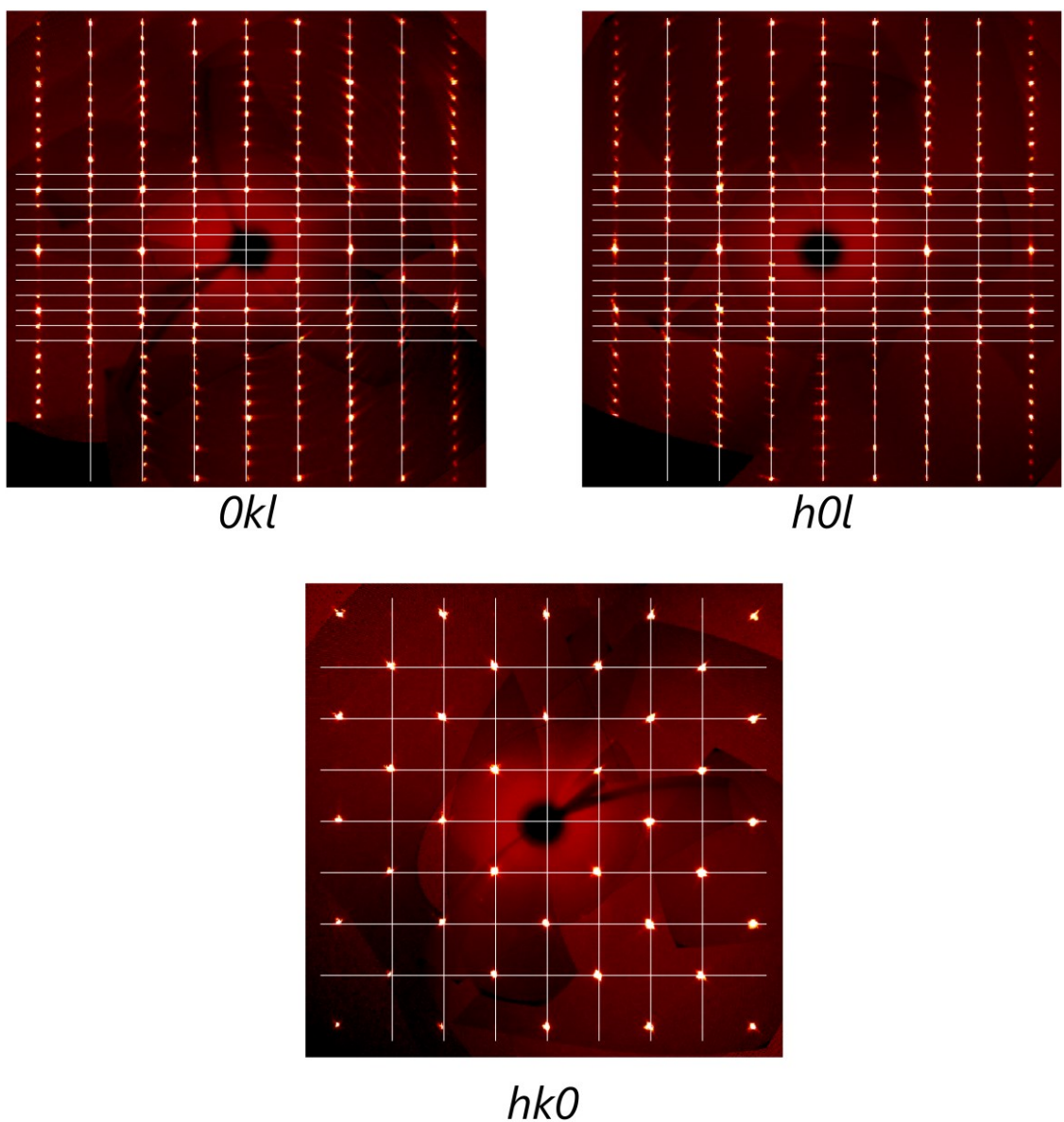

Figure 3. Precession images synthesized from the CCD single crystal diffraction data for La(O,F)BiSe$_2$. All the lines are guide for eyes.

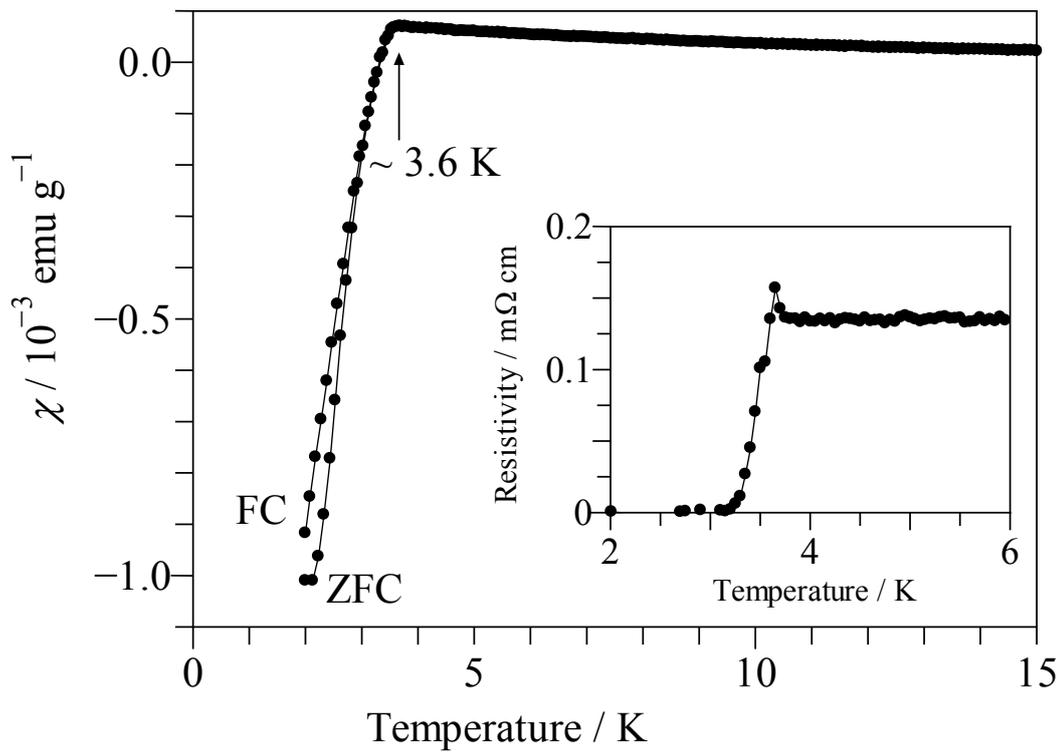

Figure 4. Temperature dependence of magnetic susceptibility of the single crystal of La(O,F)BiSe$_2$ in zero-field cooling (ZFC) and field cooling (FC) mode. The inset shows temperature dependence of resistivity with an enlargement scale of superconducting transition.